\documentclass[prl,twocolumn,letterpaper,showpacs,amsmath,amssymb,floatfix]{revtex4}
\usepackage[dvips]{graphicx}
\usepackage{amssymb}
\begin{document}

\newcommand{\ket}[1]{|#1\rangle}
\newcommand{\bra}[1]{\langle #1|}
\newcommand{\bracket}[2]{\langle #1|#2\rangle}
\newcommand{\ketbra}[1]{|#1\rangle\langle #1|}
\newcommand{\average}[1]{\langle #1\rangle}
\newtheorem{theorem}{Theorem}

\title{Quantum Key Distribution Using Equiangular Spherical Codes}
\author{Joseph M. Renes}
\affiliation{Department of Physics and Astronomy, University of New Mexico,\\
Albuquerque, New Mexico 87131--1156, USA\\
\texttt{renes@phys.unm.edu}}

\begin{abstract}
Mutually unbiased bases have been extensively studied in the literature
and are simple and effective in quantum key distribution protocols, but
they are not optimal. Here equiangular spherical codes are introduced
as a more efficient and robust resource for key distribution. Such
codes are sets of states that are as evenly spaced throughout the
vector space as possible. In the case the two parties use qubits and 
face the intercept/resend eavesdropping strategy, they can make use 
of three equally-spaced states, called a \emph{trine}, to outperform the 
original four-state BB84 protocol in both speed and reliability. 
This points toward the optimality of spherical codes in
arbitrary dimensions.
\end{abstract}


\maketitle

\noindent The possibility of secure key distribution using quantum
states is by now a well established feature of quantum information
theory.  In the original 1984 proposal of Bennett and Brassard
(BB84)~\cite{bb84}, four states of a spin-1/2 system, the eigenstates
of $\sigma_z$ and of $\sigma_x$, are used as signals by the sender
Alice. These states are naturally partitioned into two orthonormal
bases from which the receiver Bob chooses one at random to measure the
signal. Because the bases are \emph{unbiased}---i.e., the overlap
between vectors from distinct bases is always the same, equal to $1/2$
for qubits---Bob learns nothing when his measurement doesn't correspond
to Alice's preparation, but everything when it does. The
nonorthogonality of the states allows Alice and Bob to detect
eavesdropping by an adversary Eve, so the states form an
unconditionally secure cryptographic protocol~\cite{mayers98}.

One more unbiased basis, the eigenvectors of $\sigma_y$, can be added
to the BB84 set, forming a new six-state protocol~\cite{bruss98}.
Unbiased bases can be found in higher dimensions as
well~\cite{woottersfields89}, and the key distribution protocol has
been extended to such cases, with increasing dimension leading to
improved security~\cite{cbkg02}.  In these analyses, however, the
security is not proved to be unconditional, since only particular
eavesdropping attacks are studied.

Unbiased bases have been the cornerstone of key distribution schemes.
But are they optimal? For simple eavesdropping strategies, I show here
in the qubit case that they are not, suggesting that they are not
optimal for unconditional eavesdropping attacks either.  The analysis
here is based on a more efficient and robust set of states, the
\emph{equiangular spherical codes}, also known as \emph{Grassmann
frames}.  Analysis of the qubit case reveals a key distribution
protocol based on three states having equal overlap, the trine
ensemble, which is both faster and more secure than the BB84 protocol
when subjected to two simple eavesdropping attacks, intercept-resend
and cloning.  This provides compelling evidence that spherical codes
can outperform their unbiased cousins.  An analysis of spherical codes
in higher dimensions will be presented in a subsequent
paper~\cite{renes03}.



Recall the general setting of quantum key distribution. Two parties,
Alice and Bob, wish to make use of an authenticated public classical 
channel and an insecure quantum channel controlled by an adversary 
Eve to establish a secret key for the purposes of encrypting and sharing
other data. They start with a sequence of samples from a
given tripartite probability distribution shared between the three parties. 
Alice and Bob then proceed to ``distill'' the key by 
sharing information based on their individual sequences over the classical channel.
How exactly this distillation is 
achieved is an information-theoretic problem. However, for eavesdropping 
strategies in which Eve doesn't directly make use of the distillation information, 
how the probability distribution 
arises in practice and what distributions are at all possible are
purely questions of physics, and these questions are addressed in this paper.

The probability distribution arises from using the quantum channel to
send quantum information.  Alice sends quantum states drawn from a
certain signal ensemble through the channel to Bob, who performs a
specific measurement (in the case of signaling states drawn from
mutually unbiased bases, the several measurement bases Bob chooses from
for his measurement are here amalgamated into a single POVM
measurement).  Alice and Bob fix the signal ensemble and the
measurement using the public channel.  Eve is free to exploit this
information to mount an attack on their protocol, using her control of
the quantum channel; she can in principle subject the signal states to
any physical interaction that she wishes.  Alice and Bob's goal is to
exploit the quantum nature of the channel to make Eve's eavesdropping
ineffective.

The relevant probability distribution is the joint probability
$p(a_i,b_j,e_k)$ of Alice's signal, Bob's measurement result, and the
result of any measurement Eve performs in the course of eavesdropping.
Repeated use of the protocol yields a sequence of samples drawn from
this distribution.  Alice and Bob, however, must establish which
distribution they are sampling from, as it depends on Eve's attack.
Typically, Eve has some physical setup which can give rise to many
different distributions as she changes the strength of her interference
with the channel.  Given an assumption of the type of attack, Alice and
Bob determine the extent of Eve's interference by making public and
comparing a fraction of the Alice's signals and Bob's measurement
results.   Knowing the distribution $p$, they can distill a key of
length $MR$ from the remaining $M$ samples in accordance with the
bounds
\begin{equation}
\label{eq:keyratebound}
I_E\leq R\leq I(A\!:\!B|E)\;,
\end{equation}
where $I(X\!:\!Y)=H(X)+H(Y)-H(XY)$ is the mutual information of $X$ and
$Y$, $H(\cdot)$ being the Shannon entropy, and
$I_E=I(A\!:\!B)-\min\{I(A\!:\!E),I(B\!:\!E)\}$.  The lower bound
obtains when the key is distilled using one-way
communication~\cite{ck78}; to progress beyond this requires a technique
called \emph{advantage distillation}, though this is of limited
efficiency~\cite{maurer93,gisinwolf99}.

These bounds provide a method of investigating the cryptographic
usefulness of a signal ensemble.  Given a signal ensemble, Bob's
measurement, and an assumption about the nature of Eve's attack, the
probability distribution can be calculated, and the key rate bounds
determined.  In this way the security of the protocol against this
attack is established.  To say that a protocol is unconditionally
secure is to demonstrate its security against all possible attacks.

The focus now turns to Alice's signal ensemble and Bob's measurement.
An intuitively appealing ensemble is a \emph{spherical code}, a
complex-vector-space version of points on a sphere whose minimal
pairwise distance is maximal.  The complex version, called the
\emph{Grassmann packing problem}, asks for a set of unit vectors in
$\mathbb{C}^d$ whose \emph{maximal\/} pairwise overlap is
\emph{minimal}~\cite{Strohmer03}. When all these pairwise overlaps are
equal, this \emph{equiangular\/} spherical code is called a
\emph{Grassmann frame}; i.e., a set
$\mathcal{C}=\{\ket{\phi_k}\in\mathbb{C}^d\}_{k=1}^n$ for $n\!\geq\! d$
is a Grassmann frame if
\begin{equation}
\label{eq:sc}
|\bracket{\phi_j}{\phi_k}|^2=\frac{n-d}{d(n-1)}\qquad \forall\,\,
j\!\neq\! k\;.
\end{equation}

Grassmann frames also arise as the solution to the ``minimum energy
problem.''  For a set of unit vectors $\mathcal{C}$, call
$V_t(\mathcal{C})=\sum_{j,k}|\bracket{\phi_j}{\phi_k}|^{2t}$ the $t$-th
``potential energy'' of the set of the vectors~\cite{benedettofickus03}.  
The minimum energy
problem is to find $\mathcal{C}$ having $n\!\geq\! d$ elements such
that $V_1=n^2/d$ and $V_2$ is minimized.  Note that $n^2/d$ is the
global minimum of $V_1$.
This follows from considering the (at most) $d$ nonzero (real)
eigenvalues $\gamma_j$ of the Gram matrix
$G_{jk}\!=\!\bracket{\phi_j}{\phi_k}$. Clearly $\sum_k \gamma_k\!=\!n$
and $\sum_k \gamma_k^2\!=\!V_1(\mathcal{C})$.  These being the
equations for a plane and a sphere, the minimum of $V_1$ occurs if and
only if all the $\gamma_k$ are equal to $n/d$, whence $V_1$ is bounded
below by $n^2/d$.  Thus what is sought is the set of vectors with the
minimum $V_2$ energy, given minimum $V_1$ energy.

To find a lower bound for the minimum of $V_2$, let
$\lambda_{jk}=|\bracket{\phi_j}{\phi_k}|^2$, and employ the same method
again.  We have immediately that $\sum_{j\neq
k}\lambda_{jk}=V_1-n=n(n-d)/d$ and $\sum_{j\neq
k}\lambda_{jk}^2=V_2-n$, whence the minimum of $V_2$ over all sets
minimizing $V_1$ is bounded below by making all the $\lambda_{jk}$ the
same and given by Eq.~(\ref{eq:sc}).  When this lower bound is
achieved, i.e $V_2=n^2(n-2d+d^2)/(n-1)$, the result is a Grassman frame.

The existence of Grassmann frames isn't known for arbitrary $n$ and
$d$, though some general statements can be made~\cite{rbksc03}.  They always exist for
$n=d+1$ (a regular simplex), but never when $n>d^{\,2}$.  For $n\leq
d^{\,2}$, when a Grassman frame exists, it is a spherical code, but for
$n>d^{\,2}$, spherical codes aren't equiangular.

By minimizing $V_1$, Grassmann frames automatically form measurement
POVMs, which can be used by Bob to detect Alice's signal. This is true
because $S=\sum_k \ket{\phi_k}\bra{\phi_k}=(n/d)I$, so that a POVM can
be constructed from the subnormalized projectors
$(d/n)\ketbra{\phi_k}$.  To see this, fix an orthonormal basis
$\{\ket{e_k}\}$ and consider the matrix $T_{jk}=\bracket{e_j}{\phi_k}$.
The Gram matrix can be written as $G_{jk}=(T^\dagger T)_{jk}$, while
$S_{jk}=(TT^\dagger)_{jk}$, so both have the same eigenvalues.  When
$V_1$ is minimized, these $d$ eigenvalues are all $n/d$, implying that
the vectors form a resolution of the identity.

Such sets are appealing because they are the sets that are ``least
classical'' in the following sense~\cite{fuchssasaki03a}.  Consider
using these quantum states as signals on a classical channel as
follows.  Instead of sending the quantum state, Alice performs the
associated measurement and communicates the result to Bob using a
classical channel.  Bob then prepares the associated quantum state at
his end.  The fidelity of Bob's reconstruction with the input state,
averaged over inputs and measurement results, measures how well the
classical channel can be used to transmit quantum information.  This
fidelity is $dV_2/n^2$, so among all ensembles which themselves 
form POVMs, Grassmann frames are hardest to transmit
``cheaply'' in this way.  Eavesdropping on the communication between
Alice and Bob makes the channel more classical---Eve is essentially
trying to copy the signal---so one might expect that Grassmann frames
are useful in foiling the eavesdropper.

For the case of qubits, there are only two equiangular spherical codes,
the trine and the tetrahedron. These are named after their Bloch-sphere
representation: the trine is a set of three equally-spaced coplanar
vectors, and the tetrahedron is the familiar regular simplex in three
dimensions.  Here we use the following representation of the trine
states:
\begin{equation}
\ket{\phi_j}
=\frac{e^{2\pi ij/3}}{\sqrt{2}}\left(\ket{0}+e^{2\pi ij/3}\ket{1}\right)\;,
\quad
j=0,1,2.
\end{equation}

The task now is to determine key rate bounds for the trine protocol and
to compare with the original BB84 scheme.  Generically, Bob uses the
\emph{same\/} Grassmann frame to measure as Alice uses to signal.  Such
a measurement attempts to \emph{confirm\/} which state Alice sent.  For
qubits, however, Bob can construct an ``inverted measurement'' from the
states $\ket{\widetilde{\phi}_j}$ that are orthogonal to the trine
states; this measurement attempts to \emph{exclude\/} one of the
possible signal states.  By so doing, he increases the mutual
information of his outcomes with Alice's signals, thus improving the
prospects for creating a key.  This strategy doesn't work in higher
dimensions, as the orthogonal complement of a signal state isn't a pure
state.

Two eavesdropping attacks are considered here, the cloning attack and
the intercept-resend attack.  Both are single-system,
\emph{incoherent\/} attacks, as opposed to the most general
many-system, coherent attacks.  One feature of generic attacks is Eve's
ability to control the interaction strength of her probe with the
signal.  To mimic this feature in these schemes, Eve intercepts only a
fraction $q$ of the signals, allowing the rest to pass unmolested.

Now consider the two attacks in turn. The cloning attack is simple: Eve
attempts to clone the incoming signal state as best she can and then
makes the same measurement as Bob on her probe.  She implements the
unitary operator $U$ acting on the signal and her probe state,
initially in the state $\ket{0}$, which maximizes the average fidelity
$
\sum_j|\bra{\phi_j,\phi_j}U\ket{\phi_j,0}|^2/n,
$
subject to the constraint that all states are cloned equally well.  If
Eve's attack is not symmetric in this sense, Alice and Bob might be
able to improve their detection by exploiting the asymmetry.  The
resulting distribution for a clone attack is simply
\begin{equation}
p(a_i,b_j,e_k)
=\frac{4}{27}|\bra{\widetilde{\phi}_j,\widetilde{\phi}_k}U\ket{\phi_i,0}|^2\;.
\end{equation}
To describe varying $q$, Eve's random variable includes an additional
value which occurs when she does not implement $U$, in which case the
expression for the distribution is the same, but with $I$ replacing
$U$. A numerical maximization of $U$ for the trine and the four states
of the BB84 protocol was carried out using Mathematica's implementation
of the simulated annealing algorithm.  The trine result is
\begin{equation}
U=\frac{1}{\sqrt{2}}\left(
\begin{array}{cccc}
0 & 0 & 0 & \sqrt{2} \\
1 & 1 & 0 & 0 \\
1 & -1 & 0 & 0 \\
0 & 0 & \sqrt{2} & 0
\end{array}
\right)\;,
\end{equation}
for a fidelity of $(1+\sqrt{2})/4$.  For BB84, the $\pm 1$ $\sigma_z$
eigenstates are cloned to
\begin{equation}
\frac{1}{4}\left((2\!\pm\!\sqrt{2})\ket{00}\!\mp\!i\sqrt{2}(\ket{01}\!+\!\ket{10})
\!+\!(2\!\mp\!\sqrt{2})\ket{11}\right)\;,
\end{equation}
and the other two cloned states are obtained by the positive and
negative superpositions of these states. Somewhat surprisingly, the
cloning fidelity for BB84 is the same as for the trine.  More
surprisingly, \emph{cloning is useless to Eve in both cases}, since the lower
key rate bound is positive for all values of $q$.  
Cloning every signal provides Eve as much
information as Bob about Alice's string, as the cloning procedure turns out two 
copies of equal quality. However, Alice's information about Bob's string is 
still greater than Eve's, so they may use that string as the starting point 
for key distillation. 
By computing the bounds from equation~\ref{eq:keyratebound}
it is easily verified that the trine ensemble offers higher key generation rates, 
but as cloning is a very weak attack, this conclusion is of little force.

The focus now shifts to the intercept-resend attack.  This is similar
to splicing a classical channel into a quantum channel, as described
above.  Eve receives Alice's signal, measures it, creates a new quantum
state based on that measurement, and sends it on to Bob.  Due to the
symmetry of inversion between Alice and Bob's states 
it's best for Eve to include in her measurement \emph{both} ensembles. 
This ensures that her mutual information with Alice is the same as with Bob. 
Upon observing a particular result, she simply leaves the system in the corresponding state; 
thus the joint distribution when $q=1$ is quite simple:
\begin{equation}
p(a_i,b_j,e_k)=\frac{2}{27}\left\{
\begin{array}{lc}
|\bracket{\phi_i}{\widetilde{\phi}_k}|^2|\bracket{\widetilde{\phi}_k}{\widetilde{\phi}_j}|^2 &
0\leq k\leq 2\\
|\bracket{\phi_i}{\phi_k}|^2|\bracket{\phi_k}{\widetilde{\phi}_j}|^2 & k\leq 3\leq 5
\end{array}
\right.
\end{equation}
Again, for varying $q$, the probability distribution simply
includes an extra value that occurs when Eve doesn't intercept the
signal.  From this distribution it is easy to calculate the key rate
bounds and the rate $E$ of additional errors, due to Eve's attack,
which Alice and Bob observe when comparing samples using the public
channel. Considering the key rate $R$ as a function of the error rate
$E$ enables a comparison with the same quantities derived from the BB84
protocol. Figure~1 shows the upper and lower key generation rate bounds
for the trine and BB84 protocols as a function of error rate. By having
one fewer outcome, the trine is inherently better at information
transfer between the parties.  The lower bound, which is more relevant
for realistic implementation, shows that the trine is also much more
secure, tolerating roughly 9\% error.

\begin{figure}[h]
\label{fig:keyrates}
\begin{center}
\includegraphics[scale=.85]{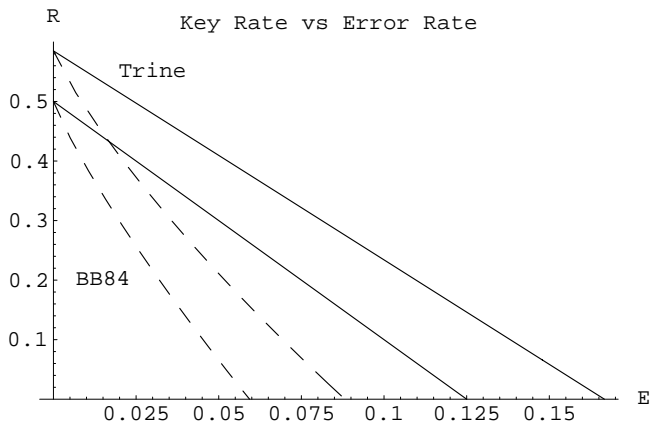}
\caption{Upper (solid) and lower (dashed) key rate bounds as a function
of error rate for the trine-based and BB84 protocols subject to the
intercept-resend attack.  For each protocol, the bounds emanate from
the same point on the vertical axis at zero error (no eavesdropping)
and drop down to zero key rate at the largest tolerable error rate on
the horizontal axis.  The trine is both faster (higher key generation
rate) and more robust (higher tolerable error) than the BB84 protocol.}
\end{center}
\end{figure}

Note that in this analysis, the usual first step in the BB84 protocol,
i.e., sifting over the public channel to determine when Bob's
measurement basis matches Alice's signal basis, cannot be performed for
the trine, as there is nothing like different bases.  Strictly
speaking, sifting belongs to the key distillation phase of the
protocol, so it is appropriate to exclude it here.

This analysis strongly suggests that the trine-based protocol might be
much more useful for key distribution than BB84, but this conclusion is
not firm, as the two attacks considered are insufficiently general.  It
is known, however, for the BB84 protocol that the intercept-resend
attack is nearly optimal~\cite{fggnp97}, so it is quite reasonable
to expect the analysis here to be indicative of the more general case.


Recently a strong relationship between secure key distribution and
entanglement has been identified by considering a coherent version of
these ``prepare-and-measure'' protocols.  Instead of preparing a state
and sending it to Bob for measurement, Alice prepares a bipartite
state, ostensibly entangled, and sends half to Bob. Each party then
measures his or her half, returning the protocol to the original
picture.  In this setting both the upper and lower key generation rate
bounds can be translated into questions of entanglement and
nonlocality.  From the upper bound, it follows that secure key
distribution is possible if the corresponding coherent
process leaves Alice and Bob with a state which is one-copy
distillable~\cite{amg03,cll03}.  From the lower bound, it follows that
key distribution is possible if the bipartite state violates some Bell
inequality~\cite{agmc03}.

Equiangular spherical codes fit nicely into this picture, as they can
always be realized from maximally entangled states.  Thus they start on
the same footing as unbiased bases, for which this is also true. To
demonstrate this, consider a spherical code
$\mathcal{C}=\{\ket{\phi_k}\}$ and a ``conjugate'' code
$\mathcal{C}^*=\{\ket{\phi_k^*}\}$ formed by complex conjugating each
code state in the standard basis.  Then it is a simple matter to show
that $\ket{\Phi}=(\sqrt{d}/n)\sum_k\ket{\phi_k}\ket{\phi^*_k}$ is
maximally entangled.  Thus if Alice prepares this state and sends the
second half to Bob, they can realize the ``prepare-and-measure'' scheme
by measurement.

The performance of the trine-based protocol establishes the usefulness
and suggests the superiority of Grassmann frames for key distribution.
Extending the intercept-resend analysis to higher dimensions is simple,
if tedious, and is done in detail elsewhere~\cite{renes03}.  The result
is the same: in every dimension, a suitable Grassmann frame can be
found to outperform the unbiased bases in both speed and reliability.

Physics dictates the distributions that can be realized, and
information theory determines how to distill a key from the data drawn
from the distribution.  It is important to remember that distillation
is relatively straightforward when using unbiased bases.  After making
many measurements, Alice and Bob sift the data to determine in which
cases they have selected the same basis.  Absent any eavesdropper, this
creates a key, and if errors are present they can employ a simple privacy amplification
scheme to ensure security.  First, they determine the error rate of
Bob's data, and knowing this, they create a secret key that Eve has
vanishingly small probability of knowing simply by taking the
EXCLUSIVE-OR of large blocks of the data.  When using equiangular
spherical codes, no such simple key distillation protocol is available.

The author acknowledges helpful input from C.~M.~Caves, A.~J.~Scott,
and K.~K.~Manne. This work was supported in part by Office of Naval
Research Grant No.~N00014-00-1-0578.


\begin{thebibliography}{99}
\bibitem{bb84}
C.~H.~Bennett and G.~Brassard, in \emph{Proceedings of the IEEE
International Conference on Computers, Systems, and Signal Processing}
(IEEE, New York, 1984), p.~175.

\bibitem{mayers98}D. Mayers, \texttt{quant-ph/9802025}.

\bibitem{bruss98}
D.~Bru\ss,
Phys. Rev. Lett. {\bf 81}, 3018 (1998).

\bibitem{woottersfields89}
W.~K.~Wootters and B.~D.~Fields,
Ann.\ Phys. (N.Y.) {\bf 191}, 363 (1989).


\bibitem{cbkg02}
N.~J.~Cerf, M.~Bourennane, A.~Karlsson, and N.~Gisin, Phys. Rev. Lett. {\bf 88}, 127902 (2002).

\bibitem{renes03}
J.~M.~Renes, ``Spherical Codes in Quantum Information Theory'', in
preparation.

\bibitem{ck78}
I.~Csisz\'{a}r and J.~K\"{o}rner,
IEEE Trans. Inf. Theory, {\bf IT-24}, 339 (1978).

\bibitem{maurer93}
U.~M.~Maurer, IEEE Trans. Inf. Th. {\bf 39}, 733 (1993).

\bibitem{gisinwolf99}
N.~Gisin and S.~Wolf, Phys. Rev. Lett. {\bf 83}, 4200 (1999).

\bibitem{Strohmer03}
T.~Strohmer and R.~Heath,
Appl.\ Comp.\ Harm.\ Anal.\ {\bf 14}, 257 (2003).

\bibitem{benedettofickus03}
J.~J.~Benedetto and M.~Fickus, 
Adv.~Comput.~Math. {\bf 18}, 357 (2003).


\bibitem{rbksc03}
J.~M.~Renes, R.~Blume-Kohout, A.~J.~Scott, and C.~M.~Caves, \texttt{quant-ph/0310075}.

\bibitem{fuchssasaki03a}
C.~A. Fuchs and M.~Sasaki,
Quant.\ Info.\ Comp.\ {\bf 3}, 377 (2003).

\bibitem{fggnp97}
C.~A.~Fuchs \emph{et~al.}, Phys. Rev. A {\bf 56}, 1163 (1997).



\bibitem{amg03}
A.~Ac\'in, L.~Masanes, and N.~Gisin,
Phys. Rev. Lett. {\bf 91}, 167901 (2003).


\bibitem{cll03}
M.~Curty, M.~Lewenstein, and N.~L\"utkenhaus,
\texttt{quant-ph/0307151}.


\bibitem{agmc03}
A.~Ac\'in, N.~Gisin, L.~Masanes, and V.~Scarani,
\texttt{quant-ph/0310166}.

\end{thebibliography}
\end{document}